# Symmetry of the Hyperfine and Quadrupole Interactions of Boron Vacancies in a Hexagonal Boron Nitride

Irina N. Gracheva,* Fadis F. Murzakhanov, Georgy V. Mamin, Margarita A. Sadovnikova, Bulat F. Gabbasov, Evgeniy N. Mokhov, and Marat R. Gafurov

**ABSTRACT:** The concept of optically addressable spin states of deep-level defects in wide band gap materials is successfully applied for the development of quantum technologies. Recently discovered negatively charged boron vacancy defects ($V_B^-$) in hexagonal boron nitride (hBN) potentially allow a transfer of this concept onto atomic-thin layers due to the van der Waals nature of the defect host. Here, we experimentally explore all terms of the $V_B^-$ spin Hamiltonian reflecting interactions with the three nearest nitrogen atoms by means of conventional electron spin resonance and high frequency (94 GHz) electron–nuclear double resonance. We establish symmetry, anisotropy, and principal values of the corresponding hyperfine interaction (HFI) and nuclear quadrupole interaction (NQI). The HFI can be expressed in the axially symmetric form as $A_\perp = 45.5 \pm 0.9$ MHz and $A_\parallel = 87 \pm 0.5$ MHz, while the NQI is characterized by quadrupole coupling constant $C_q = 1.96 \pm 0.05$ MHz with slight rhombisity parameter $\eta = (P_{xx} - P_{yy})/P_{zz} = -0.070 \pm 0.005$. Utilizing a conventional approach based on a linear combination of atomic orbitals and HFI values measured here, we reveal that almost all spin density ($\approx 84\%$) of the $V_B^-$ electron spin is localized on the three nearest nitrogen atoms. Our findings serve as valuable spectroscopic data and direct experimental demonstration of the $V_B^-$ spin localization in a single two-dimensional BN layer.

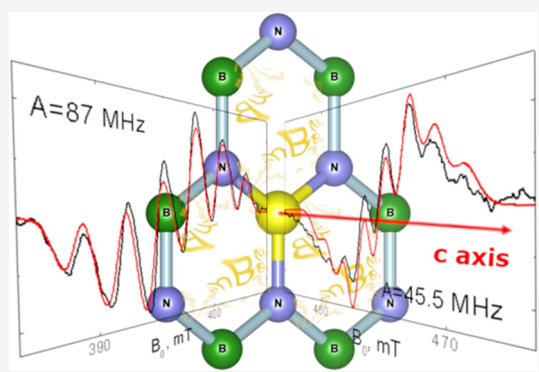

## ■ INTRODUCTION

Optically polarized electron spins of defects in solids serve as a valuable platform for development of advanced quantum technologies[1−5] and are widely used as a tool to probe unconventional many body quantum physics in condensed matter.[6,7] The development of solid-state technologies based on defects with a significant electron−nuclear interaction has been underway for several decades and has been implemented into diverse semiconductor crystal structures.[8] The main explored solid-state platforms with these respects are diamond with negatively charged nitrogen vacancy defects (NV⁻) and silicon carbide (SiC) with NV⁻ and vacancy-related defects.[1−11] Quantum magnetometers,[2,5,12] nanoscale magnetic resonance imaging (MRI),[13] sensors of thermal and electrical fields,[3,5] and masers operated at room temperature[14,15] have been proposed and developed using advantages of these defects. The main approach of their use in the above-mentioned applications is that the high spin state ($S \geq 1$) of the defect is already split in zero magnetic field and can be initialized, manipulated, and subsequently read out by optical or radio frequency means utilizing principles of electron spin resonance (ESR) and optically detected magnetic resonance (ODMR).[16] However, both diamond and SiC are three-dimensional (3D) crystals that prevent advanced nanoscale fabrication processing of these materials.

A distinctly new platform for potential realization of the above-mentioned scenarios has been recognized only recently by demonstration of optically addressable spin states of defects in two-dimensional van der Waals (vdW) materials, namely, hexagonal boron nitride (hBN).[17−21] hBN is formed by 2D atomic layers of sp²-hybridized nitrogen−boron atoms that are coupled through weak vdW interactions. Its ultrawide band gap ($E_g \approx 6$ eV)[22,23] naturally grants the existence of deep-level defects with optical transitions well below its band gap.[24] Together with a well-developed exfoliation technique allowing one to achieve fine tuning of the number of atomically thin layers (down to the single layer), these make such kind of layered material particularly interesting for nanoscale quantum technologies[23−25] and as a monolayer single-photon quantum emitter.[26] Particularly, exploiting 2D nature of hBN, defects emitting single photons have been successfully isolated in a single monoatomic BN layer,[25] and the search for defects in hBN from the perspectives of their applicability for quantum sensing and qubits has been launched.[18−21,27−29]

Among the large variety of defects possessing ODMR, only one defect in the form of a negatively charged boron vacancy ($V_B^-$) has been shown to be controllably and reproducibly generated in the hBN host[18,28,30−32] and its microscopic origin is well understood and established by means of rigorous ESR spectroscopy and calculations based on the density functional theory.[18,28,33] The structure of the $V_B^-$ defect is schematically shown in Figure 1a as a missing boron atom having three

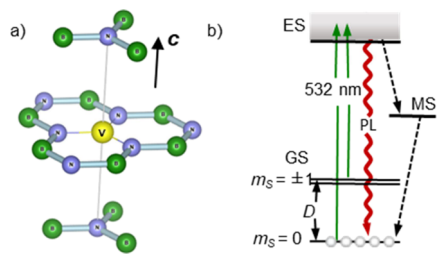

**Figure 1.** (a) Structure of the hBN lattice and schematic representation of $V_B^-$ defects, where green balls represent boron atoms, blue represents nitrogen atoms, and yellow represents the boron vacancy center; (b) $V_B^-$ center energy level diagram in the absence of external static magnetic field and the scheme of the optical pumping cycle of the ground state (GS) $m_S = 0$ spin sublevel. Excitation (green) transfers the system into the excited state (ES). Radiative recombination (purple) and spin-dependent nonradiative intersystem crossing decay to the GS via the metastable state (MS) (dashed lines). $D$ denotes the ZFS.

nitrogen atoms as the nearest neighbors.[18,28,33] The basic spin properties of the defect are summarized in Figure 1b. The spin-triplet ($S = 1$) ground state of the defect is already split in zero magnetic field ($D \cong 3.6$ GHz) giving rise to the corresponding energy separation of the $m_S = 0$ and $m_S = \pm 1$ spin sublevels known as the zero-field splitting (ZFS). Under optical excitation, the $m_S = 0$ ground-state sublevel of the defect is predominantly populated through a spin-dependent recombination channel in its optical excitation–recombination cycle, giving rise to the ability to address and readout the spin state of the defect via ODMR or ESR.[18,28,33] These serve as a background to use $V_B^-$ defects as a spin probe for quantum sensing[34,35] as well as it uses as a qubit.[18,27−29]

One of the major advantages inherently given to the negatively charged boron vacancy in hBN through its high $D_{3h}$ symmetry is that the spin density strongly bound to the defect site seems to be fully localized in a single two-dimensional layer of $sp^2$-hybridized boron and nitrogen atoms. The latter peculiarity preserves the spin properties of this defect, keeping them almost unchanged for the bulk crystal and for the single BN layer. This fact is well documented through ab initio calculations;[28,33] however, it has not been rigorously probed experimentally. Driving on the mentioned above, we focused here on the study of the anisotropy of hyperfine and nuclear quadrupole interactions associated with $V_B^-$ electron spin and three nearest nitrogen nuclei ($^{14}$N, $I = 1$) by means of ESR and pulsed electron–nuclear double resonance (ENDOR) techniques. Special attention is also paid to the study of the HFI because of its importance in determining the nature of the point defects in hBN and the effect on the coherent properties of the boron vacancy.[36] Based on the results of these experiments, we accomplish the study of the electron–nuclear interactions of the $V_B^-$ electron spin with the nearest nitrogen shell and are able to demonstrate that the $V_B^-$ spin is localized in a single two-dimensional layer.

## MATERIALS AND METHODS

The sample used in this study is commercially produced hexagonal boron nitride single crystal (HQ Graphene company) with dimensions 1 mm × 1 mm × 0.15 mm. To create boron vacancies, the sample was irradiated to a total dose of $6 \times 10^{18}$ cm$^{-2}$ with 2 MeV electrons.

The hBN crystal with mentioned characteristics was placed in a resonator system of a spectrometer by a special nonmagnetic sample holder. The ESR spectra were measured at $T = 35$ K on a standard Bruker ESR spectrometer (ESP-300) in the X-band frequency range ($\nu_{mw} \approx 9.6$ GHz), equipped with an Oxford Instruments liquid helium flow cryostat for low temperature measurements. The ENDOR spectra were measured at $T = 25$ K in the W-band frequency range ($\nu_{mw} = 94$ GHz) on a Bruker Elexsys E680 spectrometer. The spectra were recorded using a Mims pulse sequence: $\pi/2 - \tau - \pi/2 -$ radiofrequency $\pi$ pulse $- \pi/2 - \tau -$ electron spin echo (ESE), where $\pi/2 = 40$ ns, $\tau = 260$ ns, and $\pi_{rf} = 72$ μs. All ESR and ENDOR spectra were measured under optical excitation of the sample with 532 nm laser through an optical fiber. The EasySpin software package was used to simulate the obtained ESR spectra.[37]

## RESULTS AND DISCUSSION

To refine the structure of the center and in order to study the anisotropy of the electron–nuclear interactions, we first measure the stationary X-band ESR spectra of the hBN sample under $\lambda = 532$ nm optical excitation in the orientation of a static magnetic field **B** parallel (**B**∥**c**) and perpendicular (**B**⊥**c**) to the hexagonal **c** axis. Results are shown in Figure 2a. The doublet of lines labeled with vertical arrows originate from the allowed magnetic dipole transition ($m_S = 0 \leftrightarrow m_S = +1$, and $m_S = 0 \leftrightarrow m_S = -1$) between Zeeman-split triplet states. The splitting between these lines is $\Delta B \cong 257$ mT $= 2D/\gamma_e$, where $D \approx 3.6$ GHz and $\gamma_e = 28$ GHz/T is the electron gyromagnetic ratio. These values are the spectroscopic fingerprints of the $S = 1$ $V_B^-$ defect.[18,31,33] Optically induced predominant population of the $m_S = 0$ spin sublevel is readily seen through phase reversal of the fine-structure lines, as schematically depicted in the right inset in Figure 2a.

The HFI structure shown in the enlarged scale for the $m_S = 0 \rightarrow m_S = +1$ ESR line in the left inset in Figure 2a is due to the interaction of the electron spin of the defect with the nuclear spins of the nearest nitrogen atoms ($^{14}$N, nuclear spin $I = 1$, natural abundance 99.63%). To describe the ESR spectra, the following spin Hamiltonian is used.

$$H = g\mu_B \mathbf{B}\mathbf{S} + D\left(S_z^2 - \frac{S(S+1)}{3}\right) + \sum_{k=1}^{3} \mathbf{A}_k \mathbf{I}_k \mathbf{S} \quad (1)$$

Here, the first term reflects electron Zeeman interaction with isotropic g-factor, and the second term describes ZFS with the principal z-axis of the D-tensor that coincides with the **c** axis of the crystal. The third term describes the hyperfine interaction of the $V_B^-$ electron spin with three ($k = 3$) nearest to the vacancy $^{14}$N nuclear spins using the axial A-tensor with the principal z-axis directed along the nitrogen-dangling bond. As shown in Figure 2c, these directions are labeled as $A_{zz}$ for each nitrogen nuclei. $S$ and $I$ are the operators of a total electron and nuclear spins ($S = 1$, $I = 1$), $g$ is the Landé factor, $\mu_B$ is the Bohr magneton, and $B$ is the static magnetic field.

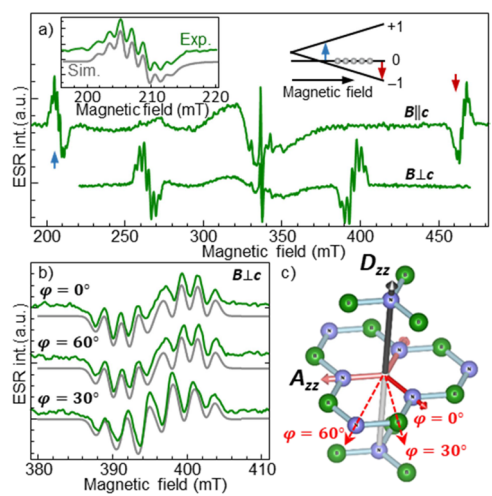

**Figure 2.** (a) ESR spectra of $V_B^-$ defects measured for $B\|c$ and $B\perp c$ orientations. The left inset shows the seven-line HFI structure of the $m_S = 0 \leftrightarrow m_S = +1$ transition for $B\|c$ in the enlarged scale, and its simulation (gray) with parameters summarized in the first row of Table 1. (b) Measured angular dependence (green) of the hyperfine structure with $B$ rotating in the 2D (0001) hexagonal plane. The corresponding simulations for three different directions of the magnetic field vector relative to the nitrogen dangling bonds are shown in gray. (c) Schematic representation of the mutual orientation of the fine splitting and hyperfine tensors' principal axes of the boron vacancy in the hBN lattice. $D_{zz}$ means the direction of the maximum value of zero field splitting which is along the crystallographic $c$-axis. Principal $z$-axes of the hyperfine interaction tensor $A$ for each nitrogen nucleus are directed along the dangling bonds, designated with bold red arrows. Schematic representation of the magnetic field $B$ rotation in the 2D-plane on certain angles $\varphi$. $\varphi = 0$ is related to the $A_{zz}$ direction.

In the $B\|c$ orientation, all three nitrogen atoms are equivalent. That is, one would expect to observe the 7-line hyperfine structure governed by the consideration that the number of the hyperfine lines is determined as $2nI + 1 = 7$, where $n$ is the number of equivalent nuclei with $I = 1$. This seven-line HF structure together with its simulation using eq 1 is shown in the left inset in Figure 2a. The extracted value of the hyperfine coupling constant $A_{xx} = 47 \pm 1$ MHz corresponds to the previously measured in the ESR/ODMR experiments[18,28] and is in good agreement with the value calculated through density functional theory (DFT).[28,33]

To determine the anisotropy of the hyperfine interaction, we then measure the angular dependence of the hyperfine splitting in the 2D plane in the $B\perp c$ configuration by rotating the $B$ vector in the (0001) plane by angle $\varphi$, as depicted in Figure 2c. We are interested in a region with a pronounced hyperfine structure in the range of magnetic fields 380–410 mT. The ESR spectra for three different angles $\varphi$ are shown in Figure 2b. A satisfactory description of the spectra is obtained for the values of the spin Hamiltonian parameters indicated in Table 1 (first row). The corresponding simulated spectra are shown in gray.

Since the stationary ESR technique does not allow one to determine parameters of the quadrupole interaction, we carry out the ENDOR experiments. An analysis of the ENDOR spectra also makes it possible to refine the values of the hyperfine interaction tensor. Like the ESR spectra, all ENDOR experiments were collected under $\lambda = 532$ nm optical excitation and for different orientations of the constant magnetic field vector $B$ relative to the $c$-axis. Figure 3a,b shows the measured ENDOR spectra for $B\perp c$ and Figure 3c for $B\|c$.

For a correct description of the ENDOR spectra, two additional terms have to be added to the spin Hamiltonian (1). Namely, the term reflecting nuclear Zeeman interaction ($Z_n = -g_n\mu_n BI$), where $g_n$ is the nuclear $g$-factor for $^{14}$N, $\mu_n$—nuclear magneton. The second term is $I_k P I_k$, where $P = \frac{3eQ_N V_{ij}}{4I(2I-1)}$, describing the nuclear quadrupole interaction (NQI). The latter is related to the interaction of the electric field gradient $V_{ij}$ with the nuclear electric quadrupole moment $Q_N$ and described using the quadrupole interaction tensor $P_{ij}$. Based on the symmetry of the system, it is logical to assume that the direction of the $z$-axis of the $P$-tensor for each nucleus coincides with the direction of the $z$-axis of the hyperfine interaction tensor (see Figure 2c).

Results of ENDOR experiments are shown in Figure 3. Dashed lines are for the simulation of the ENDOR spectra with parameters listed in Table 2. The asymmetry coefficient was taken equal to $\eta = (P_{xx} - P_{yy})/P_{zz} = -0.070$, and $C_q = 1.96$ MHz, where $C_q$ is called the quadrupole coupling constant and defined as $C_q = eQ_N V_{zz}$. The hyperfine interaction tensor values were slightly corrected to $A_{xx} \approx A_{yy} = A_\perp = 45.5$ MHz and $A_{zz} = A_\| = 87$ MHz. In the orientation $B\perp c$, the angle of the in-plane deviation of $B$ from the dangling bond according to the simulation is 7.7° degrees. It should be noted that the $C_q$ value determined here is in good agreement with previously derived $C_q = 2.11$ MHz based on the results of the electron spin echo envelop modulation (ESEEM) experiments.[28] The slight difference in 150 kHz between these two values is due to the fact that in the current experiments, the asymmetry of the NQI is revealed and thus is taken into account. The presence of additional lines in the ENDOR spectra (Figure 3a) near the Larmor frequency ($\nu_{Larmor} \approx 10.2$ MHz) of the nitrogen $^{14}$N nucleus can be caused by two reasons: (i) signals from the second nitrogen coordination sphere and (ii) signals from the nuclei of neighboring layers, from below and above. Since such hyperfine interactions from distant (remote) nuclei are weak, ENDOR signals from them are absent in Figure 3b. The corresponding weak electron–nuclear interactions manifest themselves only through quadrupole splitting at spin transitions with $m_S = 0$.

**Table 1. Spectroscopic Values of the Spin Hamiltonian Including the g-Factor, the Terms of the Fine Structure, and the Hyperfine Interaction**[a]

| $g_{iso}$ | $D$, MHz | $A_{xx}$ ($x\|c$), MHz | $A_{yy}$, MHz | $A_{zz}$, MHz | ref |
|---|---|---|---|---|---|
| 2.001(1) | 3600 ± 10 | 47 ± 1 | 47 ± 1 | 85 ± 1 | in this work |
| 2.000 | 3600 | 47 | | | 18 |
| | 3470 | 48 | | | 36 |
| ab initio analysis | | | | | |
| | 3470 | 44.97 | 46.12 | 87.15 | 28 |
| | 3467 | 46.110 | 47.935 | 91.571 | 33 |

[a]The table also presents previously measured[18,36] and calculated[28,33] parameters of the $V_B^-$ spin-Hamiltonian for comparative analysis.

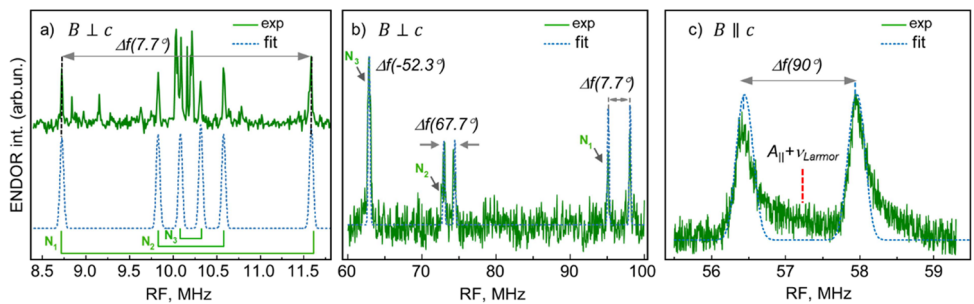

**Figure 3.** Electron−nuclear interaction spectra (green) revealed by ENDOR together with corresponding simulations (blue). (a) Nuclear magnetic resonance (NMR) transitions in the $^{14}$N Larmor frequency region corresponding to the nuclear spin flips in the $m_S = 0$ manifold. Three pairs of lines, labeled as $N_1$, $N_2$, and $N_3$, reflect NQI with three nonequivalent nitrogen nuclei; (b) NMR transitions related to three nonequivalent nitrogen nuclei $N_1$, $N_2$, and $N_3$ relative to the external magnetic field; (c) ENDOR spectrum in the **B**∥$c$ orientation at which all three nitrogen nuclei are equivalent, i.e., with the same values of the hyperfine coupling constant $A_{xx}$. Dashed blue lines are calculated ENDOR transitions using eq 1 with additional terms of NQI and nuclear Zeeman interaction.

**Table 2. Values of the Hyperfine and Quadrupole Interaction Tensors Determined by Simulating the ENDOR Spectra**

| | hyperfine interaction (MHz) | | | quadrupole interaction (MHz) | | | |
|---|---|---|---|---|---|---|---|
| | $A_{xx}$ ($x$∥$c$) | $A_{yy}$ | $A_{zz}$ | $P_{xx}$ ($x$∥$c$) | $P_{yy}$ | $P_{zz}$ | $C_q = eQ_N V_{zz}$ |
| $^{14}$N | 45.5 ± 0.9 | 45.5 ± 0.9 | 87 ± 0.5 | −0.52 ± 0.02 | −0.46 ± 0.02 | 0.98 ± 0.03 | 1.96 ± 0.05 |

General description of the ENDOR frequencies arising due to the quadrupole splitting can be obtained by the following expression $\Delta f(\varphi) \approx \frac{3}{4}C_q(3\cos^2\varphi - 1)$, taking into account all rotational transformations in Euler angles (see Figure 2c). Here, $\varphi$ is the angle between external magnetic field **B** and the nitrogen dangling bonds, $C_q = 1.96$ MHz. This is why we observe the ENDOR signal separately from each nitrogen nucleus, as labeled in Figure 3 as $N_1$, $N_2$, and $N_3$, i.e., in this case, each HFI value depends on the angle as $A_{iso} + T(3\cos^2(\varphi) - 1)$, where $A_{iso}$ and $T$ are the isotropic part and the anisotropic part of HFI, correspondingly.

Up to this point, we show that the $A$-tensor has an axial symmetry about the local $z$-axis, which corresponds to the direction of the dangling bonds (the direction of the $z$-axis of the $A$-tensor is 90° with the $z$-axis of the $D$-tensor). It is also seen that the anisotropy has an axial symmetry of the 6th order, which corresponds to perfectly coordinated three nitrogen atoms closest to the vacancy at the vertices of a regular triangle. Using refined values of hyperfine interaction constants from ENDOR data as $A_\perp = 45.5$ MHz and $A_\parallel = 87$ MHz, it is possible to determine the isotropic ($A_{iso}$) and anisotropic ($T$) parts of the HFI: $A_{iso} = \frac{2A_\perp + A_\parallel}{3} \approx 59.3$ MHz and $T = \frac{A_\parallel - A_\perp}{3} \approx 13.8$ MHz. The isotropic part of the HFI induced by the Fermi contact interaction is a measure of the spin density at the nitrogen nucleus (2s character of the wave function). The anisotropic part induced by the dipole−dipole interaction corresponds to the spin density concentrated on the 2p nitrogen orbital. Both parts of the HFI are represented as follows:

$$A_{iso} = \frac{8\pi}{3}g\mu_B g_N \mu_N c_{2s}^2 \eta_j^2 |\psi_{2s}(0)|^2$$

$$T = \frac{2}{5}g\mu_B g_N \mu_N c_{2p}^2 \eta_j^2 \langle r^{-3} \rangle_{2p} \quad (2)$$

Here, $\eta_j^2$ is the electron spin density, $\psi_{2s}(0)$ is the 2s wave function at the sight of nitrogen nuclei, and $\langle r^{-3} \rangle_{2p}$ is the averaging over the 2p electronic wave function. Representing the wave function of an unpaired electron as a linear combination of atomic orbitals $\Psi = \sum_j \eta_j \psi_j$ and $\psi_j = c_{2s}\psi_{2s} + c_{2p}\psi_{2p}$, where $\psi_{2s}$ and $\psi_{2p}$ correspond to 2s and 2p orbitals and $c_{2s}$ and $c_{2p}$ are corresponding coefficients, it is possible to calculate the electron spin density $\eta_j^2$ [38,39] localized on the nitrogen atom through the known values of $\psi_{2s}(0)$ and $\langle r^{-3} \rangle_{2p}$ for nitrogen. In this case, the normalization condition $c_{2s}^2 + c_{2p}^2 = 1$ is taken into account and it is assumed that the atomic orbitals of nitrogen in hBN do not differ much from those previously calculated for the case of a free nitrogen atom.[40] Thus, the calculated spin density on one nitrogen atom is $\eta_j^2 = 28\%$. Since there are three equivalent nitrogen atoms in the nearest environment of a boron vacancy, the spin density concentrated on them is ≈84%. The extraction of the hyperfine Fermi contact term and the calculation of the electron spin density within the monolayer play an important role in the coherent properties of the boron vacancy.[36]

## ■ CONCLUSIONS

In this work, the features of electron−nuclear interactions of the boron vacancy $V_B^-$ with surrounding equivalent nitrogen nuclei were investigated by the methods of the stationary electron paramagnetic resonance (9.6 GHz) and high-frequency (94 GHz) pulsed ENDOR technique. In addition to the $g$-factor and fine splitting ($D = 3.6$ GHz), the detailed study of the angular dependence of the ESR spectra made it possible to determine all the components of the hyperfine interaction tensor ($A_\perp = 45.5 ± 0.9$ MHz and $A_\parallel = 87 ± 0.5$ MHz) and therefore to establish its axial symmetry. The established HFI value serves as an estimate of $V_B^-$ electron spin density localization in a two-dimensional plane of $sp^2$-hybridized BN atoms. Given the isotropic $A_{iso}$ and anisotropic $T$ contributions of HFI, we can conclude that nearly all spin density of the $V_B^-$ density (≈84%) is localized on the three nitrogen atoms in the 2D BN layer. The latter means that the number of layers in a single crystal will not have a significant effect on the ground-state spin properties of the center, which means that it might be possible to obtain a monoatomic hBN layer with optically addressable structurally protected spin sublevels of color centers.

The ENDOR method applied at two different canonical orientations allows us to determine the value of the quadrupole interaction $C_q$ = 1.96 ± 0.05 MHz with an estimate of the asymmetry parameter $\eta$ = −0.070 ± 0.005. Despite the fact that the fine structure tensor is directed along the crystallographic axis $c$, the quadrupole interaction has its maximum value along the dangling bond. The relatively small value of the asymmetry parameter $\eta$ indicates the axial symmetry of the quadrupole interaction. Hence, in this work, we explored all terms of the $V_B^-$ spin Hamiltonian together with their symmetry reflecting interactions with the three nearest nitrogen atoms.

■ AUTHOR INFORMATION

**Corresponding Author**

Irina N. Gracheva − *Institute of Physics, Kazan Federal University, Kazan 420008, Russia*; orcid.org/0000-0003-3715-8170; Email: subirina@gmail.com

**Authors**

Fadis F. Murzakhanov − *Institute of Physics, Kazan Federal University, Kazan 420008, Russia*; orcid.org/0000-0001-7601-6314

Georgy V. Mamin − *Institute of Physics, Kazan Federal University, Kazan 420008, Russia*; orcid.org/0000-0002-7852-917X

Margarita A. Sadovnikova − *Institute of Physics, Kazan Federal University, Kazan 420008, Russia*; orcid.org/0000-0002-5255-9020

Bulat F. Gabbasov − *Institute of Physics, Kazan Federal University, Kazan 420008, Russia*; orcid.org/0000-0003-1359-3717

Evgeniy N. Mokhov − *Ioffe Institute, St. Petersburg 194021, Russia*

Marat R. Gafurov − *Institute of Physics, Kazan Federal University, Kazan 420008, Russia*; orcid.org/0000-0002-2179-2823

Complete contact information is available at:
https://pubs.acs.org/10.1021/acs.jpcc.2c08716

**Author Contributions**

The manuscript was written through contributions of all authors.

**Funding**

This research was funded by the RSF grant No. 20-72-10068.

**Notes**

The authors declare no competing financial interest.

■ ACKNOWLEDGMENTS

The authors would like to thank Russian Science Foundation grant no. 20-72-10068 for supporting the study. Irina N. Gracheva, Fadis F. Murzakhanov, and Georgy V. Mamin thank Victor Soltamov for participating in the discussion and interpretation of the results.

■ ABBREVIATIONS

| | |
|---|---|
| hBN | hexagonal boron nitride |
| ESR | electron spin resonance |
| ENDOR | electron nuclear double resonance |
| ODMR | optically detected magnetic resonance |
| ZFS | zero field splitting |
| HFI | hyperfine interaction |
| ESE | electron spin echo |
| NMR | nuclear magnetic resonance |
| NQI | nuclear quadrupole interaction |